\documentstyle[twoside,fleqn,epsf,proceedings]{article}
\def\bc{\begin{center}}
\def\ec{\end{center}}
\def\be{\begin{equation}}
\def\ee{\end{equation}}
\newcommand{\labar}{\overline{\Lambda}}

\newcommand{\kkinetic}{\bar{Q}(x)\, \vec{D}^{2}\, Q(x)}
\newcommand{\kkrinetic}{\bar{Q}(x)\, \vec{D}^{2}_{R}\, Q(x)}
\newcommand{\energy}{\bar{Q}(x)\, D_{4}\, Q(x)}
\newcommand{\unity}{\bar{Q}(x)\, Q(x)}

\newcommand{\beq}{\begin{equation}}
\newcommand{\eeq}{\end{equation}}
\newcommand{\beqn}{\begin{eqnarray}}
\newcommand{\eeqn}{\end{eqnarray}}

\newcommand{\AmS}{{\protect\the\textfont2
  A\kern-.1667em\lower.5ex\hbox{M}\kern-.125emS}}

\title{Renormalons on the Lattice}

\author{M.~Crisafulli$^{\rm a}$,
V.~Gim\'enez\address{I.N.F.N. and Dipartimento di Fisica dell' Universit\'a
di Roma ``La Sapienza'' \\ Piazza Aldo Moro 2, I-00185 Roma, Italy},
G.~Martinelli$^{\rm a}$\address{Theory Division, CERN, 1211 Geneva 23,
Switzerland}\thanks{Talk presented by G.~Martinelli},
C.T.~Sachrajda\address{Dept. of Physics,  University of  Southampton,
Southampton SO17 1BJ, UK.}}
\begin{document}
\begin{abstract}
We present the first lattice calculation of the B-meson binding energy
$\labar$  and of the kinetic energy $\lambda_1/2 m_Q$ of the heavy-quark inside
the pseudoscalar  B-meson. In order to cancel the ambiguities
due to the ultraviolet renormalons present in the operator
matrix elements,  this calculation has required
 the non-perturbative subtraction of the
power divergences present  in  the Lagrangian operator  $\energy$ and
in the kinetic energy operator $\kkinetic$. The non-perturbative
renormalization of the relevant  operators
has been implemented by imposing suitable renormalization
 conditions on quark matrix elements in the Landau gauge.
\end{abstract}
\maketitle
\section{Introduction} \label{intro}
Among the quantities which cannot be predicted on the
basis of the Heavy Quark Effective Theory
(HQET) \cite{Neubert}  there are several  parameters which characterize the
dynamics of strong interactions,
such as the heavy quark binding energy $\labar$,
relevant for  higher order corrections
to the semileptonic form factors, and the
heavy quark kinetic energy $\lambda_1/2m_Q$,
which enters in the  predictions of many inclusive
decay rates.
Lattice  HQET offers the
possibility of a numerical, non\- -per\-tur\-ba\-ti\-ve  determination
of these quantities
 from first principles and without free parameters.

The parameter  $\labar$ denotes the asymptotic  value of the difference
between the hadron and the heavy quark ``pole" mass $m_Q$
\be \labar\, =\, \lim_{m_{Q} \rightarrow \infty}\, \left( M_{H}\, -\, m_{Q}
\right) .
\label{eq:naivedef}
\ee
It has been recently shown that the pole mass is ambiguous due to the
presence of infrared renormalon singularities  \cite{beneke}. At lowest
order in $1/m_Q$, the infrared renormalon ambiguity
 appearing in the definition of the pole mass is closely related to
the ultra-violet renormalon singularity present in the matrix elements of
the operator $\energy$. This singularity is due to the linear  power divergence
  of  $\energy$,  induced by  its mixing
with the lower dimensional operator $\bar Q(x) Q(x)$.
 On the lattice  the linear divergence
manifests itself as a factor proportional
to  the inverse lattice spacing $1/a$
in the  mixing coefficient of the operator $\bar Q(x) Q(x)$. In
ref.  \cite{mms}, it was stressed that these divergences must be subtracted
non-perturbatively since factors such as
\be \frac{1}{a}\exp \Bigl( - \int^{g_0(a)} \frac{d g^\prime}{\beta(g^\prime)}
\Bigr) \sim
 \Lambda_{QCD}, \label{eq:argu} \ee
which do not appear in perturbation theory,  give non-vanishing contributions
as $a \to 0$.
 Renormalons represent
an explicit example of non-perturbative effects of this kind.

The matrix elements of the kinetic energy operator
also contain power divergent contributions. In this case, the
origin of the divergences is the mixing of
 $\kkinetic$ with the operator $\energy$, with a
coefficient that diverges linearly, and with the scalar density
 $\unity$, with a quadratically divergent
coefficient  \cite{mms}.

Following the non-perturbative method for eliminating
the power divergences proposed in ref. \cite{d2reno},
we have computed the ``physical" values of $\labar$
and $\lambda_1$. This method will be explained in sec. \ref{labardef}.

By fixing the non-perturbative renormalization conditions
of  $\energy$ by using  the heavy quark propagator
  in the Landau gauge, we found
\be \bar \Lambda= (232 \pm 22 \pm 30 \pm 25)\,  {\rm MeV} \label{eq:vlb} \ee
where the first error is  statistical  and the others
 systematic and will be explained in sec. \ref{numerical}.

In order to remove the power divergences from the
kinetic energy operator, we have imposed  to the relevant  operator
  a renormalization condition which corresponds to the
``physical"  requirement
$\langle Q(\vec p=0)\vert \kkinetic \vert Q(\vec p=0) \rangle = 0$.
This renormalization condition has been used to extract the values of the
renormalization
constants, that have been obtained
 with a small statistical error. Unfortunately, after the  subtraction of
the power divergences, we were only able to obtain  a loose upper bound
$  \lambda_1 < 1.0\, {\rm GeV}^2$.
\section{Non-perturbative definition of $\labar$ and $\lambda_1$}
\label{labardef}

The cancellation of  the power divergences of the operators
 $\energy$ and  $\kkinetic$   is achieved by imposing
appropriate renormalization conditions on the quark matrix elements
\cite{d2reno}.
In numerical simulations, quark and gluon propagators can be computed
non-perturbatively
 by working in a fixed gauge, typically the Landau gauge
 \cite{nperenor}.
 On general grounds, we expect that the heavy quark propagator,
at lowest order in $1/m_Q$ has the form
\be
 S(x)\, =\,
\delta(\vec x)\, \theta(t)\, A(t)\, \exp (-\lambda t) ,
\label{prop}
\ee
where $ S(x)\, =\, \langle S(\vec x ,t\vert \vec 0,0) \rangle$,
\beqn S(\vec x ,t\vert \vec 0 ,0 )= \delta(\vec x )
\, \theta(t)\,  \exp\Bigl(i \int_0^t A_0(t^\prime) dt^\prime \Bigr)
\nonumber \eeqn  being the non-translational invariant
propagator for a given gauge field configuration.
 $\langle \dots \rangle$ represents the average over the gauge
field configurations and   $A(t)$ is an unknown smooth function of $t$,
such that $\ln \Bigl( A(t+a)/A(t) \Bigr) \to 0 $ as $t \to \infty$.
The constant $\lambda$
is linearly divergent in $1/a$ and  is associated with the
ultraviolet renormalon in of the   heavy-quark propagator.
We can remove it by using
\beqn
{\cal L}_{\rm eff} \,=\, \frac{1}{1+ \delta m a}
\Bigl( \bar Q(x)\, D_4\, Q(x)\, +\,
\delta m\, \bar{Q}(x)\, Q(x)\Bigr),
\label{eq:l0effp}
\nonumber \eeqn
which corresponds to the propagator
\be
 S^{ \prime}(\vec x ,t)\, =\,  \delta(\vec x)\, \theta(t)\,
A(t)  \exp{(\,- [\lambda -\delta \overline{m} ]\, t)}. \label{propp}  \ee
with
\beqn
 -\, \delta \overline{m}\, \equiv\, \frac{\ln(1+\delta m a)}{a}\, =\,
\lim_{t \to \infty} \delta \overline{m}(t)=
 \nonumber  \eeqn \beqn
 \lim_{t \to \infty}
  \frac{1}{a}\, \ln\left(\frac{S(\vec x ,t+a)} {S(\vec x ,t)}
\right) \to -\lambda\, +\, O(\frac{1}{t}) .
\label{ct}  \eeqn

We are now in a position of defining the renormalized binding energy
$\labar$
using ${\cal E}$, the bare ``binding" energy usually computed from the
two point heavy-light meson correlation functions \cite{sommer}
\beqn
C(t) = \sum_{\vec x}\, \langle 0|\,\bar Q(\vec x,t)\Gamma
q(\vec x,t)\; \bar q(\vec 0,0) \Gamma Q(\vec 0,0)\,|0\rangle
\nonumber
\eeqn
\be \rightarrow Z^2 \exp(-{\cal E}t)
\label{eq:ctbig}
\ee
Thus
\be
\labar\, \equiv\, {\cal E}\, -\, \delta \overline{m}\, ,
\label{lare}
\ee

The renormalized kinetic operator, free of power divergences
has the form
\begin{eqnarray}
\kkrinetic  = \kkinetic
- \frac{C_1}{a}  \eeqn \beqn
\Bigl(\energy\, + \delta m \unity \Bigr)
 - \frac{C_{2}}{a^{2}} \unity,
\nonumber \label{eq:d2ren}
\end{eqnarray}
where the constants $C_{1}$ and $C_{2}$ are a function
of the bare lattice coupling constant $g_0(a)$.
In order to eliminate  the quadratic and linear  power divergences,
a possible  non-perturbative  renormalization condition for
 $\kkrinetic$  is that its subtracted matrix element, computed
 for a quark at rest in the Landau gauge, vanishes
$
\langle Q(\vec p=0) \vert \kkrinetic \vert Q(\vec p=0) \rangle=0.
$
This is  equivalent to defining the subtraction constants through
the relation
\beqn
  R_{\vec{D}^{2}}(t)  \equiv  C_1\,+\,C_2\, t = \nonumber \eeqn
\beqn
\frac{ \sum^t_{\vec{x},\vec{y},t^\prime=0}\,   \langle \,
S^{\prime}(\vec x,t\vert
\vec y,t')\,  \vec D^2_y(t')\, S^{\prime}(\vec y,t'\vert
\vec 0,0)
\, \rangle}{
\sum_{\vec{x}}\,\langle S^{\prime}(\vec x,t\vert \vec 0,0\rangle
}
\label{eq:c12}
\end{eqnarray}
By fitting the time dependence of
$R_{\vec{D}^{2}}(t)$ to eq. (\ref{eq:c12}),
 one obtains $C_{1,2}$.
The relation between   the mass of the meson  and  the mass
of the quark  to order $1/m_{Q}$ is then given by
\beqn
M_{H}\, =\, m_{Q}\, +\, {\cal E}\, -\, \delta \overline{m}\, +\,
\frac{\lambda_1 - C_{2}}{2\, m_{Q}} +O(\frac{1}{m_Q^2})
\label{eq:nextor}
\eeqn
Notice that only the constant $C_{2}$ enters the eq. (\ref{eq:nextor}) because
$C_{1}$ is eliminated by using the equations
of motions.

\section{Numerical implementation of the renormalization procedure}
\label{numerical}

The non-perturbative, numerical renormalization of $\energy$ and $\kkinetic$
has been performed  by using the heavy quark propagators and matrix elements
computed on a statistical sample of 36 gluon
configurations, generated by numerical simulation
 on a $16^{3}\times 32$ lattice at $\beta=6.0$.  The heavy-light
meson propagators have been computed
 using the
improved SW-Clover  action  \cite{clover} for the light quarks,
in the quenched approximation.   For the binding
energy $\cal {E}$ we  made use of the high statistics
 results obtained by the APE collaboration at $\beta=6.0$,
using the Wilson \cite{alltonw} and the SW-Clover action \cite{allton}.
\begin{figure}
\begin{center}
\begin{picture}(160,100)
\put(-70,-120){\special{gim2.ps}}
\end{picture}
\end{center}
\caption{Effective mass of the heavy-quark propagator $S_{H}(t)$ as a
function of the time. The curve represents a  fit of the numerical results
(in the improved case) to the expression given in eq. (\ref{eq:fit1}).}
\label{fig:fit}
\end{figure}

In fig. \ref{fig:fit},  we present the values of $\delta \overline{m}(t)$
 as a function of time.  Inspired by
one loop perturbation theory at small values of $a/t$,
 we made a fit to $\delta \overline{m}(t)$
using the expression
\be a \, \delta \overline{m}(t)  = a \, \delta \overline{m} +
\gamma\frac{a}{t} , \label{eq:fit1} \ee
where $\delta \overline{m}$ and $\gamma$ are the parameters of
the fit.
We have also used different expressions
to fit $\delta \overline{m}(t)$
and changed the interval of the fits in order to check the stability
of the value of the results.
Our best estimate of the mass counter-term is
\be a \,\delta \overline{m} = 0.50 \pm 0.01 \pm 0.02 \label{dbm} \ee
where the first error is statistical and the second the systematic one from
the different extrapolation procedures.
In order to evaluate $\labar$, we have used the results of
the high statistics calculations of ${\cal E}$ given in refs.
 \cite{alltonw,allton}. They obtained  $a \, {\cal E}_W=0.600(4)$.
with the Wilson action and $a \, {\cal E}_{SW}=0.616(4)$
in the Clover case. The difference between the results obtained with
different actions
${\cal E}_{SW}-{\cal E}_W=(0.616-0.600)\, a^{-1} \sim 30$ MeV give us
a conservative
 estimate of $O(a)$ effects in the determination of this quantity.

We are now ready to present our prediction for $\labar$.
Using $\delta \overline{m}$ from eq. (\ref{dbm}) and the SW-Clover
determination of ${\cal E}$, we  quote
\be \labar = (232 \pm  22 \pm 30 \pm 25)\,  {\rm MeV}, \label{final} \ee
where the first error is the statistical one, the second is
our   estimate of $O(a)$ effects and the third comes from  the
calibration of the value of the lattice spacing.
\begin{figure}
\begin{center}
\begin{picture}(160,100)
\put(-85,-110){\special{gim3.ps}}
\end{picture}
\end{center}
\caption{The ratio $R_{\vec{D}^{2}}(t)$ as a function of the time.
The linear fit  is also given.}
\label{fig:c1c2t}
\end{figure}

In fig. \ref{fig:c1c2t}, we plot $R_{\vec{D}^{2}}(t)$, as defined in
eq. (\ref{eq:c12}), as a function of the time $t$. The numerical
results are in remarkable agreement with the predicted
linear behaviour. We notice that the constant $C_2$ is obtained,
 with a modest sample
of configurations, with a precision of  $\sim 5\%$.
In order to compute $\lambda_1$, we have also
computed  the three-point correlation
function
\beqn
\sum_{\vec x, \vec y}\, \langle 0|\,
J(\vec x ,t)\, \left[\, \bar Q(\vec y,t') \vec{D}^{2}_{y} Q(\vec y,t')\,
\right]
\, J^\dagger(\vec 0,0) \,|0 \rangle=
\nonumber
\eeqn
\beqn
C_{\vec{D}^{2}}(t,t') \rightarrow Z^2\, \lambda_1\,
\exp(-({\cal E} - \delta \overline{m} )\, t)
\eeqn
for sufficiently large euclidean time
distances  $t'$ and $| t - t'|$.
Therefore, we can determine $\lambda_1$ by taking the ratio
\be
R(t,t')\, =\, \frac{C_{\vec{D}^{2}}(t,t')}{C(t)}\, \rightarrow\, \lambda_1
\ee
as usually done in numerical simulations.
We have obtained
  the unrenormalized value  $a^2 \, \lambda_1=-0.75 \pm 0.15$,
and hence
\be a^2 \, \lambda_1 - C_2 =0.06 \pm 0.15 .\ee
{}From the above result, we can at most put a loose upper bound
$\vert  \lambda_1\, -\, \frac{C_{2}}{a^2} \vert \, <\, 1.0\, {\rm GeV}^{2}
$

We have shown that lattice numerical simulations give the opportunity
of defining unambiguously the important phenomenological
parameters $\labar$ and $\lambda_1$. By matching the full to the effective
 theory, this will allow more accurate theoretical
predictions of quantities relevant in heavy flavour physics.

\end{document}